\documentstyle[12pt]{article}
\makeatletter
 
 \@addtoreset{equation}{section}
\makeatother

\newcommand{\labeleq}[1]{\label{eqn:{#1}}}
\newcommand{\refeq}[1]{\ref{eqn:{#1}}}

\newcommand{\bfig}{\begin{figure}}
\newcommand{\efig}{\end{figure}}

\newcommand{\beq}{\begin{equation}}
\newcommand{\eeq}{\end{equation}}
\newcommand{\beqn}{\begin{eqnarray}}
\newcommand{\eeqn}{\end{eqnarray}}

\def\del{\partial}
\def\bfr{{\bf r}}
\def\bfv{{\bf v}}
\def\bfw{{\bf w}}
\def\bfW{{\bf W}}
\def\bfwt{\tilde{{\bf w}}}
\def\Ga{\Gamma}
\def\ga{\gamma}
\def\de{\delta}
\def\ep{\epsilon}
\def\la{\lambda}
\def\si{\sigma}
\def\slash#1{\setbox0=\hbox{$#1$}#1\hskip-\wd0\hbox to\wd0{\hss\sl/\/\hss}}
\def\Tr{{\rm Tr}}

\begin{document}
\begin{titlepage}
\begin{flushright}
       {\normalsize  OU-HET 307 \\  hep-th/0003069\\
        March, 2000 }
\end{flushright}

\begin{center} 
  {\large \bf Anomalous Interactions of Five Dimensional \\
 $USp(2k)$ Gauge Theory}
\footnote{This work is supported in part 
 by the Grant-in-Aid  for Scientific Research (10640268) 
 the Grant-in-Aid  for Scientific Research Fund (97319)
from the Ministry of Education, Science and Culture, Japan.}

\vfill
         {\bf Y.~Arakane}  \\
            and \\
         {\bf H.~Itoyama}\\

        Department of Physics,\\
        Graduate School of Science, Osaka University,\\
        Toyonaka, Osaka, 560-0043, Japan\\
\end{center}
\vfill
\begin{abstract}
We consider the five dimensional $USp(2k)$ gauge theory which consists of one antisymmetric and
 $n_{f}$ fundamental hypermultiplets.
This gauge theory is a many-probe generalization of the $SU(2)$ gauge theory in five dimensions considered
 by Seiberg in the context of probing type $I$ superstring by a D4-brane.
This gauge theory can also be obtained from the $USp(2k)$ matrix model
 by matrix T-dual transformations in the large $k$ limit.

We exhibit the anomalous interaction associated with this five dimensional theory on the new phase,
where the vacuum expectation values ($vev$s) of the scalars belonging to
 the antisymmetric hypermultiplet are also nonvanishing.
On the Coulomb phase, the anomalous interaction has been computed in \cite{Seiberg, IMS}.
\end{abstract}
\vfill
\end{titlepage}

\section{Introduction}
In recent years, attention has been paid to properties of supersymmetric gauge theories with matter multiplets
 in various dimensions.
In particular, parity odd interactions obtained from one-loop fermionic determinant or equivalently hexagon diagrams
 have received much interest till now \cite{Tsey, Th} in their interpretation as interactions among branes~\cite{Li}.
They also give distinct contribution to the imaginary part of the effective action and make sense beyond renormalizability.
These interactions are often called anomalous interactions
 (or Wess-Zumino type terms albeit the fact that can be represented locally in $D > 4$ dimensions).

In this paper, we will provide another example of anomalous interactions.
We consider the supersymmetric $USp(2k)$ gauge theory in five dimensions with
 one matter hypermultiplet in the antisymmetric representation and
 $n_{f}$ matter hypermultiplets in the fundamental representation.
We consider this theory on the new phase where the $vev$s of the scalars belonging to
 the antisymmetric hypermultiplet are also nonvanishing.
This theory is a many-probe generalization of $SU(2)$ gauge theory with $n_{f}$ fundamental matters and
 is also related to the $USp(2k)$ matrix model~\cite{IT, ITsCMK, ITs} via matrix T duality operation.
In the former case, our result exhibits a magnetic interaction among D4-branes
 in nontrivial gauge backgrounds.
On the Coulomb phase, the anomalous interaction has been computed in \cite{Seiberg, IMS}.

In section 2, we exhibit the five dimensional SYM lagrangian dealt with in this paper and
 find the background configurations of our model.
In section 3, we present our calculation and our final result is eq.~({\refeq{result}})

\section{Set up}

\subsection{Lagrangian of Five Dimensional $USp(2k)$ Gauge Theory}
We discuss the five dimensional worldvolume gauge theory associated with $USp(2k)$ matrix model.
The lagrangian of this five dimensional theory is given by 
\beq
{\cal L} = {\cal L}_{adj} + {\cal L}_{asym} + {\cal L}_{fund},
\eeq
where
\beq
{\cal L}_{adj} =
\frac{1}{g^2} {\Tr} \left(
- \frac{1}{4} {v}_{\mu \nu} {v}^{\mu \nu}
+ \frac{1}{2} [{\cal D}_{\mu}, {v}_{7}] [{\cal D}^{\mu}, {v}_{7}]
+ \frac{i}{2} {\bar{\Psi}}_{(adj)} {\Ga}^{\mu} [{\cal D}_{\mu} , \Psi_{(adj)}]
- \frac{1}{2} {\bar{\Psi}}_{(adj)} {\Ga}^{7} [{v}_{7} , \Psi_{(adj)}]
\right),
\eeq

\beqn
{\cal L}_{asym}
&=& \frac{1}{g^2} {\Tr} \left(
\sum_{M_{+} = 5, 6, 8, 9} \frac{1}{2} [{\cal D}_{\mu} , {v}_{M_{+}}] [{\cal D}^{\mu} , {v}_{M_{+}}]
+ \sum_{M_{+} = 5, 6, 8, 9} \frac{1}{2} [{v}_{7} , {v}_{M_{+}}] [{v}_{7} , {v}_{M_{+}}] \right) \nonumber \\
&+& \frac{1}{g^2} {\Tr} \left(
\frac{i}{2} {\bar{\Psi}}_{(asym)} {\Ga}^{\mu} [{\cal D}_{\mu} , \Psi_{(asym)}]
- \frac{1}{2} {\bar{\Psi}}_{(asym)} {\Ga}^{7} [{v}_{7} , \Psi_{(asym)}] \right) \nonumber \\
&-& \frac{1}{g^2} {\Tr} \left(
\sum_{M_{+} = 5, 6, 8, 9} \frac{1}{2} {\bar{\Psi}}_{(asym)} {\Ga}^{M_{+}} [{v}_{M_{+}} , \Psi_{(adj)}]
+ \sum_{M_{+} = 5, 6, 8, 9} \frac{1}{2} {\bar{\Psi}}_{(adj)} {\Ga}^{M_{+}} [{v}_{M_{+}} , \Psi_{(asym)}] \right) \nonumber \\
&+& \frac{1}{g^2} {\Tr} \sum_{M_{+}, N_{+} = 5, 6, 8, 9} \frac{1}{4} [{v}_{M_{+}} , {v}_{N_{+}}] [{v}_{M_{+}} , {v}_{N_{+}}],
\eeqn

\beqn
{\cal L}_{fund}
&=& \frac{1}{g^2} \sum_{f=1}^{n_{f}} \left(
\sum_{M_{+} = 5, 6, 8, 9} \frac{1}{2} {\cal D}_{\mu} {v}_{(f)}{}_{M_{+}} \cdot {\cal D}^{\mu} {v}_{(f)}{}_{M_{+}}
+ \sum_{M_{+} = 5, 6, 8, 9} \frac{1}{2} {v}_{7} {v}_{(f)}{}_{M_{+}} \cdot {v}_{7} {v}_{(f)}{}_{M_{+}} \right) \nonumber \\
&+& \frac{1}{g^2} \sum_{f=1}^{n_{f}} \left(
\frac{i}{2} {\bar{\Psi}}_{(f)} {\Ga}^{\mu} {\cal D}_{\mu} \Psi_{(f)}
- \frac{1}{2} {\bar{\Psi}}_{(f)} {\Ga}^{7} {v}_{7} \Psi_{(f)} \right) \nonumber \\
&-& \frac{1}{g^2} \sum_{f=1}^{n_{f}} \left(
\sum_{M_{+} = 5, 6, 8, 9} \frac{1}{2} m_{f}{}^{2} {v}_{(f)}{}_{M_{+}} \cdot {v}_{(f)}{}_{M_{+}}
+ \frac{1}{2} m_{f} {\bar{\Psi}}_{(f)} \Ga^{7} \Psi_{(f)} \right) \nonumber \\
&-& \frac{1}{g^2} \sum_{f=1}^{n_{f}} \left(
\sum_{M_{+} = 5, 6, 8, 9} \frac{1}{2} {\bar{\Psi}}_{(f)} {\Ga}^{M_{+}} \Psi_{(adj)} {v}_{(f)}{}_{M_{+}}
+ \sum_{M_{+} = 5, 6, 8, 9} \frac{1}{2} {v}_{(f)}{}_{M_{+}} \cdot {\bar{\Psi}}_{(adj)} {\Ga}^{M_{+}} \Psi_{(f)} \right) \nonumber \\
&+& \frac{1}{g^2} \sum_{f=1}^{n_{f}} \frac{1}{4} \left( \sum_{M_{+} = 5, 6, 8, 9} {v}_{(f)}^{2}{}_{M_{+}} \right)^{2}.
\eeqn
Here ${v}_{\mu}$ is the five dimensional gauge field, and
 ${v}_{7}$, ${v}_{M_{+}}$, and ${v}_{(f)}$ are respectively $USp(2k)$ adjoint, antisymmetric, and fundamental scalars.
$\Psi_{(adj)}$, $\Psi_{(asym)}$, and $\Psi_{(f)}$ are respectively $USp(2k)$ adjoint, antisymmetric, and fundamental fermions.
These fermions can be represented, using thirty-two component Majorana-Weyl spinors in 9+1 dimensions,
 which satisfy $C \bar{\Psi}^{t} = \Psi$, $\Ga_{11} \Psi = \Psi$,
\beq
\ga \Psi_{(adj)} = \Psi_{(adj)}, \ \ga \Psi_{(asym)} = - \Psi_{(asym)}, \ \ga \Psi_{(f)} = - \Psi_{(f)},
\eeq
where $\ga \equiv \Ga^{5} \Ga^{6} \Ga^{8} \Ga^{9}$.

Let us pause for a moment to discuss that this five dimensional lagrangian can be understood from the action of type IIB matrix model~\cite{IKKT},
 followed by the $USp$ projections~\cite{IT}.
\beq
S({\underline{v}}_{M} , \underline{\Psi}) =
\frac{1}{g^2} {\Tr} \left( \frac{1}{4} [ {\underline{v}}_{M} , {\underline{v}}_{N} ] [ {\underline{v}}^{M} , {\underline{v}}^{N} ] 
- \frac{1}{2} {\bar{\underline{\Psi}}}{\Ga}^{M} [ {\underline{v}}_{M} , \underline{\Psi} ] \right),
\eeq
where ${\underline{v}}_{M}$ are bosonic coordinates, and $\underline{\Psi}$ is a thirty-two component Majorana-Weyl spinor,
 which satisfies $C \bar{\underline{\Psi}}^{t} = \underline{\Psi}$, $\Ga_{11} \underline{\Psi} = \underline{\Psi}$.
 All underlined fields are in $u(2k)$-valued.
We can obtain the action of $USp(2k)$ matrix model by introducing the projectors ${\hat{\rho}}_{b \mp}$, ${\hat{\rho}}_{f \mp}$ 
which act on underlined fields,
\beq
S({\hat{\rho}}_{b \mp} {\underline{v}}_{M} , {\hat{\rho}}_{f \mp} {\underline{\Psi}})
+ \Delta S,
\eeq
where
\beqn
{\hat{\rho}}_{b \mp} {\underline{v}}_{M_{-}} &=& v_{M_{-}} = \sum_{a} v_{M_{-}}^{a} T_{a}, \\
{\hat{\rho}}_{b \mp} {\underline{v}}_{M_{+}} &=& v_{M_{+}} = \sum_{a} v_{M_{+}}^{a} X_{a}, \\
{\hat{\rho}}_{f \mp} \underline{\Psi} &=&
\sum_{a} \frac{1}{2} (1 + \ga) \Psi^{a} T_{a} + \sum_{a} \frac{1}{2} (1 - \ga) \Psi^{a} X_{a} \nonumber \\
&=& \sum_{a} \Psi_{(adj)}^{a} T_{a} + \sum_{a} \Psi_{(asym)}^{a} X_{a}.
\eeqn
Here $T_{a}$ and $X_{a}$ are, respectively, adjoint and antisymmetric representation matrices of $USp(2k)$. 
$M_{-} = 0, 1, 2, 3, 4, 7$, $M_{+} = 5, 6, 8, 9$.
We find that
 $S({\hat{\rho}}_{b \mp} {\underline{v}}_{M} , {\hat{\rho}}_{f \mp} {\underline{\Psi}}) 
= S_{adj} + S_{asym} \equiv S_{adj + asym}$ 
is the reduced action of $d=4$, ${\cal N}=2$ super Yang-Mills with one antisymmetric matter.
$\Delta S$ contains $S_{fund}$,
 which is the zero dimensional reduced action of $n_{f}$ ${\cal N}=2$ fundamental matters in $d=4$.
The part in $\Delta S$ which is not contained in $S_{fund}$ is irrelevant to
the rest of our discussion.
For more detail, see ref.~\cite{ITs}.
We obtain the lagrangian of the five dimensional gauge theory via matrix T-dual transformation 
with respect to $x^{0}$, $x^{1}$, $\ldots$, $x^{4}$ directions,
 or replacement $i v_{\mu}$ with the covariant derivative ${\cal D}_{\mu} = \del_{\mu} + i v_{\mu}$
 for $\mu = 0, 1, \ldots, 4$.

For the purpose of our calculation, we would like to regard the present five dimensional lagrangian
 as the reduction of higher dimensional one,
 when we compute the anomalous interaction~\cite{Th}. 
On the Coulomb phase, we can consider $S_{adj}$, $S_{asym}$ and $S_{fund}$ as the reductions of $d=6$, ${\cal N}=1$ supersymmetric theories. 
However, on the new phase where the $vev$s of the scalars belonging to
 the antisymmetric hypermultiplet are also nonvanishing,
 it is easier to regard $S_{adj + asym}$ as the reduction,
 with projections, of $d=10$, ${\cal N}=1$ supersymmetric theory.

\subsection{Vacuum Solutions}
We will compute the anomalous interaction on the new phase in the next section. 
Here we set all fermionic backgrounds to zero and
 we find the background configurations of our model.
From the equations of motion for bosonic fields, 
\beqn
[ {v}_{\mu} , {v}_{7} ] &=& 0 , \nonumber \\
{[}{v}_{M_{-}} , {v}_{N_{+}}] &=& 0 , \nonumber \\
v_{(f)}{}_{M_{+}} &=& 0 .
\eeqn
We find 
\beqn
v_{7} &=& diag (v_{7}^{(1)} , v_{7}^{(2)} , \ldots , v_{7}^{(k)} , -v_{7}^{(1)} , -v_{7}^{(2)} , \cdots , -v_{7}^{(k)}), \\
v_{M_{+}} &=& diag (v_{M_{+}}^{(1)} , v_{M_{+}}^{(2)} , \ldots , v_{M_{+}}^{(k)} , v_{M_{+}}^{(1)} , v_{M_{+}}^{(2)} , \ldots , v_{M_{+}}^{(k)}),
\eeqn
and all the fundamental bosonic fields $v_{(f)}$ vanish. 
The gauge field is in Cartan subalgebra of $USp(2k)$. 

\subsection{Adjoint and Antisymmetric Representation Matrices}
We present all the elements $T$ of $usp(2k)$ Lie algebra, which satisfy $T^{t}F + FT = 0$ and $T^{\dag} = T$,
\beqn
T_{0i} &=& \si_{z} \otimes e_{ii} \ (i = 1, \ldots, k), \\
T_{1ij} &=& \si_{x} \otimes \frac{1}{\sqrt{2}} e_{\{ ij\} } \ (1 \leq i < j \leq k), \\
T_{2ij} &=& \si_{y} \otimes \frac{1}{\sqrt{2}} e_{\{ ij\} } \ (1 \leq i < j \leq k), \\
T_{3ij} &=& \si_{z} \otimes \frac{1}{\sqrt{2}} e_{\{ ij\} } \ (1 \leq i < j \leq k), \\
T_{4ij} &=& {\bf 1}_{2} \otimes \frac{-i}{\sqrt{2}} e_{[ij]} \ (1 \leq i < j \leq k), \\
T_{5ij} &=& \si_{x} \otimes e_{ii} \ (i = 1, \ldots, k), \\
T_{6ij} &=& \si_{y} \otimes e_{ii} \ (i = 1, \ldots, k),
\eeqn
where $\si_{x}$, $\si_{y}$, and $\si_{z}$ are Pauli matrices and ${\bf 1}_{2}$ is the unit matrix of size 2.
$e_{ij}$ is a $k \times k$ matrix such that the element $(e_{ij})_{kl}=\de_{ik}\de_{jl}$, 
and $e_{\{ ij\} } \equiv e_{ij} + e_{ji}$, $e_{[ij]} \equiv e_{ij} - e_{ji}$.
$T_{0i}$ is Cartan subalgebra of $usp(2k)$.
We define
\beqn
H_{e^{i}} &=& T_{0i} \ (i = 1, \ldots, k), \\
T_{\pm (e_{i} + e_{j})} &=& \frac{1}{\sqrt{2}} (T_{1ij} \pm i T_{2ij}) \ (1 \leq i < j \leq k), \\
T_{\pm (e_{i} - e_{j})} &=& \frac{1}{\sqrt{2}} (T_{3ij} \pm i T_{4ij}) \ (1 \leq i < j \leq k), \\
T_{\pm 2e_{i}} &=& \frac{1}{\sqrt{2}} (T_{5ij} \pm i T_{6ij}) \ (i = 1, \ldots, k),
\eeqn
where $e^{i}$ are $k$ dimensional basis vectors and $e_{j}$ are dual basis vectors, and $e^{i} \cdot e_{j} = \de^{i}_{j}$.
The commutation relation of Cartan subalgebra $H_{e^{i}}$ and $T_{\bfw}$ is
\beq
[H_{e^{i}} , T_{\bfw}] = e^{i} \cdot \bfw T_{\bfw},
\eeq
where $\bfw \in \bfW_{adj} \equiv \{\{ \pm (e_{i} + e_{j}), \pm (e_{i} - e_{j}), \pm 2e_{i} \}\}$ is the root vector of $USp(2k)$.

Next, we present all the elements of antisymmetric representation matrices.
The element $X$ is expressed as
\beq
X=
\left(
\begin{array}{cc}
A+iC & B-iD \\
-B-iD & A-iC
\end{array}
\right) ,
\eeq
where $A$ is a real symmetric matrix, and $B$, $C$, $D$ are real skew-symmetric matrices.
All the elements of antisymmetric representation matrices are
\beqn
X_{0i} &=& {\bf 1}_{2} \otimes e_{ii} \ (i = 1, \ldots, k), \\
X_{1ij} &=& \si_{x} \otimes \frac{-i}{\sqrt{2}} e_{[ij]} \ (1 \leq i < j \leq k), \\
X_{2ij} &=& \si_{y} \otimes \frac{-i}{\sqrt{2}} e_{[ij]} \ (1 \leq i < j \leq k), \\
X_{3ij} &=& {\bf 1}_{2} \otimes \frac{1}{\sqrt{2}} e_{\{ ij\} } \ (1 \leq i < j \leq k), \\
X_{4ij} &=& \si_{z} \otimes \frac{-i}{\sqrt{2}} e_{[ij]} \ (1 \leq i < j \leq k).
\eeqn
We define
\beqn
\tilde{H}_{e^{i}} &=& X_{0i} \ (i = 1, \ldots, k), \\
X_{\pm (e_{i} + e_{j})} &=& \frac{1}{\sqrt{2}} (X_{1ij} \pm i X_{2ij}) \ (1 \leq i < j \leq k), \\
X_{\pm (e_{i} - e_{j})} &=& \frac{1}{\sqrt{2}} (X_{3ij} \pm i X_{4ij}) \ (1 \leq i < j \leq k).
\eeqn
The commutation relation of Cartan subalgebra $H_{e^{i}}$ and $X_{\bfw}$ is
\beq
[H_{e^{i}} , X_{\bfw}] = e^{i} \cdot \bfw X_{\bfw},
\eeq
where $\bfw \in \bfW_{asym} \equiv \{\{ \pm (e_{i} + e_{j}), \pm (e_{i} - e_{j}) \}\}$ 
is the weight vector of antisymmetric representation.
Diagonal matrices $\tilde{H}_{e^{i}}$ commute with $H_{e^{i}}$.

The commutation relation of a diagonal matrix $\tilde{H}_{e_{i}}$ and $T_{\bfw}$ is
\beq
[\tilde{H}_{e^{i}} , T_{\bfw}] = e^{i} \cdot \bfwt X_{\bfw},
\eeq
where
\beqn
\bfwt = \pm ( e_{i} - e_{j} )
&& \qquad {\rm for \quad} \bfw = \pm ( e_{i} - e_{j} ), \nonumber \\
\bfwt = \pm ( e_{i} - e_{j} )
&& \qquad {\rm for \quad} \bfw = \pm ( e_{i} + e_{j} ), \nonumber \\
\bfwt = 0
&& \qquad {\rm for \quad} \bfw = \pm 2 e_{i}.
\eeqn
Similarly, the commutation relation of a diagonal matrix $\tilde{H}_{e_{i}}$ and $X_{\bfw}$ is
\beq
[\tilde{H}_{e^{i}} , X_{\bfw}] = e^{i} \cdot \bfwt T_{\bfw}.
\eeq

In terms of $H_{e_{i}}$ and $\tilde{H}_{e_{i}}$, we can express the vacuum solutions as
\beq
v_{7} = \sum_{i=1}^{k} v_{7}^{(i)} H_{e_{i}} \equiv \bfv_{7} \cdot {\bf H},
\eeq
\beq
v_{M_{+}} = \sum_{i=1}^{k} v_{M_{+}}^{(i)} \tilde{H}_{e_{i}} \equiv \bfv_{M_{+}} \cdot \tilde{{\bf H}}.
\eeq

\section{Anomalous Interactions of Five Dimensional $USp(2k)$ \\
Gauge Theory}

\subsection{Computation of the Anomalous Interactions}
We compute the anomalous interactions on the new phase where the $vev$s of the scalars belonging to
 the antisymmetric hypermultiplet are also nonvanishing.

Firstly, we compute the contribution to the one-loop effective action
 by the adjoint and antisymmetric fermions,
\beq
\Ga_{\rm 1-loop}^{adj+asym} = - \frac{i}{2} \Tr \frac{1 + \Ga_{11}}{2} \hat{\rho}_{f \mp} \ln \slash{D}, \labeleq{a1}
\eeq
where $\slash{D} \equiv \Ga^{\mu} (\del_{\mu} + i v_{\mu}) + \Ga^{7} i v_{7} + \Ga^{M_{+}} i v_{M_{+}}$.

Under the variation of the gauge field, eq.~(\refeq{a1}) is
\beqn
\de \Ga_{\rm 1-loop}^{adj+asym}
&=& \frac{1}{2} \int d^{5}x \Tr \frac{1 + \Ga_{11}}{2} \hat{\rho}_{f \mp}
\Ga^{\mu} \bfw \cdot \de \bfv_{\mu} \langle x| \frac{1}{\slash{D}} |x \rangle \nonumber \\
&=& \frac{1}{2} \int d^{5}x \Tr \frac{1 + \Ga_{11}}{2} \hat{\rho}_{f \mp}
\Ga^{\mu} \bfw \cdot \de \bfv_{\mu} \langle x| \frac{\slash{D}}{\slash{D}^{2}} |x \rangle .
\eeqn
From $\Ga^{M} \Ga^{N} = \frac{1}{2} \{ \Ga^{M}, \Ga^{N} \} + \frac{1}{2} [ \Ga^{M}, \Ga^{N} ] = \eta^{MN} + \frac{1}{2} [ \Ga^{M}, \Ga^{N} ]$, we obtain
\beqn
\slash{D}^{2}
&=& D^{\mu} D_{\mu} + (\bfw \cdot \bfv_{7})^{2} + \sum_{M_{+}} (\bfwt \cdot \bfv_{M_{+}})^{2} + i \Ga^{\mu} \Ga^{\nu} \bfw \cdot \del_{\mu} \bfv_{\nu} \nonumber \\
&+& i \Ga^{\mu} \Ga^{7} \bfw \cdot \del_{\mu} \bfv_{7} + i \Ga^{\mu} \Ga^{M_{+}} \bfwt \cdot \del_{\mu} \bfv_{M_{+}}
\eeqn
We note that all $v_{M}$ are diagonal matrices, and $[v_{M}, v_{N}]$ vanish. 

We want the terms which is proportional to the epsilon symbol and does not involve the metric,
 so we keep the contribution to the imaginary part of the one-loop effective action in the Euclidean formalism.
\beqn
\lefteqn{\de \Ga_{\rm 1-loop}^{adj+asym}} \nonumber \\
&\sim& \frac{i}{2} \int d^{5}x \Tr \frac{1 + \Ga_{11}}{2} \hat{\rho}_{f \mp}
\Ga^{\mu} \bfw \cdot \de \bfv_{\mu} (\Ga^{7} \bfw \cdot \bfv_{7} + \Ga^{M_{+}} \bfwt \cdot \bfv_{M_{+}})
\langle x| \frac{1}{\slash{D}^{2}} |x \rangle  \nonumber \\
&\sim& \frac{i}{2} \int d^{5}x \Tr \frac{1 + \Ga_{11}}{2} \hat{\rho}_{f \mp}
\Ga^{\mu} \bfw \cdot \de \bfv_{\mu} (\Ga^{7} \bfw \cdot \bfv_{7} + \Ga^{M_{+}} \bfwt \cdot \bfv_{M_{+}}) \nonumber \\
&& \times \langle x| \frac{1}{\del_{\phi}^{2} + i \Ga^{\mu} \Ga^{\nu} \bfw \cdot \del_{\mu} \bfv_{\nu}
+ i \Ga^{\mu} \Ga^{7} \bfw \cdot \del_{\mu} \bfv_{7} + i \Ga^{\mu} \Ga^{M_{+}} \bfwt \cdot \del_{\mu} \bfv_{M_{+}}} |x \rangle 
\nonumber \\
&\sim& \frac{i}{2} \sum_{\bfr = adj, asym} \sum_{\bfw \in \bfW_{\bfr}} \int d^{5}x \Tr \frac{1 + \Ga_{11}}{2} \frac{1 + (-)^{|{\bfr}|} \ga}{2}
\Ga^{\mu} \bfw \cdot \de \bfv_{\mu} ( \Ga^{7} \bfw \cdot \bfv_{7} + \Ga^{M_{+}} \bfwt \cdot \bfv_{M_{+}}) \nonumber \\
&& \times \sum_{n=0}^{\infty} \langle x| \frac{(-i)^{n}
( \Ga^{\nu} \Ga^{\la} \bfw \cdot \del_{\nu} \bfv_{\la} + \Ga^{\nu} \Ga^{7} \bfw \cdot \del_{\nu} \bfv_{7}
+ \Ga^{\nu} \Ga^{M_{+}} \bfwt \cdot \del_{\nu} \bfv_{M_{+}})^{n}}{(\del_{\phi}^{2})^{n+1}} |x \rangle  \nonumber \\
&\sim& \frac{i}{8} \sum_{\bfr = adj, asym} (-)^{|\bfr|} \sum_{\bfw \in \bfW_{\bfr}} \int d^{5}x
\Tr \Ga_{11} \ga \Ga^{\mu} \bfw \cdot \de \bfv_{\mu} \Ga^{7} \bfw \cdot \bfv_{7}
\langle x| \frac{(-i)^{2} ( \Ga^{\nu} \Ga^{\la} \bfw \cdot \del_{\nu} \bfv_{\la})^{2}}{(\del_{\phi}^{2})^{3}} |x \rangle  \nonumber \\
&+& \frac{i}{8} \sum_{\bfr = adj, asym} \sum_{\bfw \in \bfW_{\bfr}} \int d^{5}x
\Tr \Ga_{11} \Ga^{\mu} \bfw \cdot \de \bfv_{\mu} \Ga^{7} \bfw \cdot \bfv_{7}
\langle x| \frac{(-i)^{4} ( \Ga^{\nu} \Ga^{M_{+}} \bfwt \cdot \del_{\nu} \bfv_{M_{+}})^{4}}{(\del_{\phi}^{2})^{5}} |x \rangle  \nonumber \\
&+& \frac{i}{8} \sum_{\bfr = adj, asym} \sum_{\bfw \in \bfW_{\bfr}} \int d^{5}x
\Tr \Ga_{11} \Ga^{\mu} \bfw \cdot \de \bfv_{\mu} \Ga^{M_{+}} \bfwt \cdot \bfv_{M_{+}} \nonumber \\
&& \times \langle x| \frac{4 (-i)^{4} ( \Ga^{\nu} \Ga^{7} \bfw \cdot \del_{\nu} \bfv_{7}) ( \Ga^{\la} \Ga^{M_{+}} \bfwt \cdot \del_{\la} \bfv_{M_{+}})^{3}}{(\del_{\phi}^{2})^{5}} |x \rangle  \nonumber \\
&=& -4i \sum_{\bfr = adj, asym} (-)^{|\bfr|} \sum_{\bfw \in \bfW_{\bfr}} \int d^{5}x
\bfw \cdot \bfv_{7} \ep^{\mu \nu \la \rho \si} \bfw \cdot \de \bfv_{\mu} (\bfw \cdot \del_{\nu}) (\bfv_{\la} \bfw \cdot \del_{\rho} \bfv_{\si})
\langle x| \frac{1}{(\del_{\phi}^{2})^{3}} |x \rangle  \nonumber \\
&+& 4i \sum_{\bfr = adj, asym} \sum_{\bfw \in \bfW_{\bfr}} \int d^{5}x
\bfw \cdot \bfv_{7} \ep^{\mu \nu \la \rho \si} \bfw \cdot \de \bfv_{\mu}
(\bfwt \cdot \del_{\nu} \bfv_{M_{+}}) (\bfwt \cdot \del_{\la} \bfv_{N_{+}}) (\bfwt \cdot \del_{\rho} \bfv_{P_{+}}) (\bfwt \cdot \del_{\si} \bfv_{Q_{+}}) \nonumber \\
&& \times \ep^{M_{+} N_{+} P_{+} Q_{+}} \langle x| \frac{1}{(\del_{\phi}^{2})^{5}} |x \rangle  \nonumber \\
&+& 16i \sum_{\bfr = adj, asym} \sum_{\bfw \in \bfW_{\bfr}} \int d^{5}x
\ep^{\mu \nu \la \rho \si} \bfw \cdot \de \bfv_{\mu} \bfw \cdot \del_{\nu} \bfv_{7}
(\bfwt \cdot \del_{\la} \bfv_{N_{+}}) (\bfwt \cdot \del_{\rho} \bfv_{P_{+}}) (\bfwt \cdot \del_{\si} \bfv_{Q_{+}}) (\bfwt \cdot \bfv_{M_{+}}) \nonumber \\
&& \times \ep^{M_{+} N_{+} P_{+} Q_{+}} \langle x| \frac{1}{(\del_{\phi}^{2})^{5}} |x \rangle , \nonumber \\
\labeleq{a5}
\eeqn
where 
\beq
\del_{\phi}^{2} \equiv \del_{\mu} \del^{\mu} + (\bfw \cdot \bfv_{7})^{2} + \sum_{M_{+}} (\bfwt \cdot \bfv_{M_{+}})^{2},
\eeq
and $(-)^{|{\bfr}|}=1$ for $\bfr = adj$, $(-)^{|{\bfr}|}=-1$ for $\bfr = asym$.
The value of $\langle x| (\del_{\phi}^{-2})^{n} |x \rangle $ is given by
\beq
\langle x| \frac{1}{(\del_{\phi}^{2})^{n}} |x \rangle 
=\frac{i}{(2 \sqrt{\pi})^{5}} \frac{\Ga (n - \frac{5}{2})}{\Ga (n)}
\left[ (\bfw \cdot \bfv_{7})^{2} + \sum_{M_{+}} (\bfwt \cdot \bfv_{M_{+}})^{2} \right]^{\frac{5}{2} - n}.
\eeq

We substitute this equation into eq.~(\refeq{a5}),
\beqn
\lefteqn{\de \Ga_{\rm 1-loop}^{adj+asym}} \nonumber \\
&=& \frac{1}{16 \pi^{2}} \sum_{\bfr = adj, asym} (-)^{|\bfr|} \sum_{\bfw \in \bfW_{\bfr}} \int d^{5}x \left[ (\bfw \cdot \bfv_{7})^{2} + \sum_{M_{+}} (\bfwt \cdot \bfv_{M_{+}})^{2} \right]^{- \frac{1}{2}} \bfw \cdot \bfv_{7} \nonumber \\
&& \times \ep^{\mu \nu \la \rho \si} \bfw \cdot \de \bfv_{\mu} (\bfw \cdot \del_{\nu} \bfv_{\la}) (\bfw \cdot \del_{\rho} \bfv_{\si}) \nonumber \\
&+& \frac{-1}{256 \pi^{2}} \sum_{\bfr = adj, asym} \sum_{\bfw \in \bfW_{\bfr}} \int d^{5}x \left[ (\bfw \cdot \bfv_{7})^{2} + \sum_{M_{+}} (\bfwt \cdot \bfv_{M_{+}})^{2} \right]^{- \frac{5}{2}} \bfw \cdot \bfv_{7} \nonumber \\
&& \times \ep^{\mu \nu \la \rho \si} \bfw \cdot \de \bfv_{\mu}
\bfwt \cdot \del_{\nu} \bfv_{M_{+}} (\bfwt \cdot \del_{\la} \bfv_{N_{+}}) (\bfwt \cdot \del_{\rho} \bfv_{P_{+}}) (\bfwt \cdot \del_{\si} \bfv_{Q_{+}}) \ep^{M_{+} N_{+} P_{+} Q_{+}} \nonumber \\
&+& \frac{-1}{64 \pi^{2}} \sum_{\bfr = adj, asym} \sum_{\bfw \in \bfW_{\bfr}} \int d^{5}x \left[ (\bfw \cdot \bfv_{7})^{2} + \sum_{M_{+}} (\bfwt \cdot \bfv_{M_{+}})^{2} \right]^{- \frac{5}{2}} \bfwt \cdot \bfv_{M_{+}} \nonumber \\
&& \times \ep^{\mu \nu \la \rho \si} \bfw \cdot \de \bfv_{\mu}
(\bfw \cdot \del_{\nu} \bfv_{7}) (\bfwt \cdot \del_{\la} \bfv_{N_{+}}) (\bfwt \cdot \del_{\rho} \bfv_{P_{+}}) (\bfwt \cdot \del_{\si} \bfv_{Q_{+}}) \ep^{M_{+} N_{+} P_{+} Q_{+}}. \nonumber \\
\labeleq{a3}
\eeqn
We can simplify the first term in eq.~(\refeq{a3}),
\beqn
\lefteqn{\de \Ga_{\rm 1-loop}^{adj+asym}} \nonumber \\
&=& \frac{1}{16 \pi^{2}} \sum_{\bfw \in \{\{ \pm 2e_{i} \}\} } \int d^{5}x \left[ (\bfw \cdot \bfv_{7})^{2} \right]^{- \frac{1}{2}} \bfw \cdot \bfv_{7}
\ep^{\mu \nu \la \rho \si} \bfw \cdot \de \bfv_{\mu} (\bfw \cdot \del_{\nu} \bfv_{\la}) (\bfw \cdot \del_{\rho} \bfv_{\si}) + \ldots \nonumber \\
&=& \frac{1}{16 \pi^{2}} \sum_{\bfw \in \{\{ \pm 2e_{i} \}\} } \int d^{5}x sgn(\bfw \cdot \bfv_{7})
\ep^{\mu \nu \la \rho \si} \bfw \cdot \de \bfv_{\mu} (\bfw \cdot \del_{\nu} \bfv_{\la}) (\bfw \cdot \del_{\rho} \bfv_{\si}) + \ldots. \nonumber \\
\eeqn

Finally, we obtain
\beqn
\lefteqn{\Ga_{\rm 1-loop}^{adj+asym}} \nonumber \\
&=& \frac{1}{48 \pi^{2}} \sum_{\bfw \in \{\{ \pm 2e_{i} \}\} } \int d^{5}x sgn(\bfw \cdot \bfv_{7})
\ep^{\mu \nu \la \rho \si} \bfw \cdot \bfv_{\mu} (\bfw \cdot \del_{\nu} \bfv_{\la}) (\bfw \cdot \del_{\rho} \bfv_{\si}) \nonumber \\
&+& \frac{-1}{256 \pi^{2}} \sum_{\bfr = adj, asym} \sum_{\bfw \in \bfW_{\bfr}} \int d^{5}x \left[ (\bfw \cdot \bfv_{7})^{2} + \sum_{M_{+}} (\bfwt \cdot \bfv_{M_{+}})^{2} \right]^{- \frac{5}{2}} \bfw \cdot \bfv_{7} \nonumber \\
&& \times \ep^{\mu \nu \la \rho \si} \bfw \cdot \bfv_{\mu}
(\bfwt \cdot \del_{\nu} \bfv_{M_{+}}) (\bfwt \cdot \del_{\la} \bfv_{N_{+}}) (\bfwt \cdot \del_{\rho} \bfv_{P_{+}}) (\bfwt \cdot \del_{\si} \bfv_{Q_{+}}) \ep^{M_{+} N_{+} P_{+} Q_{+}} \nonumber \\
&+& \frac{-1}{64 \pi^{2}} \sum_{\bfr = adj, asym} \sum_{\bfw \in \bfW_{\bfr}} \int d^{5}x \left[ (\bfw \cdot \bfv_{7})^{2} + \sum_{M_{+}} (\bfwt \cdot \bfv_{M_{+}})^{2} \right]^{- \frac{5}{2}} \bfwt \cdot \bfv_{M_{+}} \nonumber \\
&& \times \ep^{\mu \nu \la \rho \si} \bfw \cdot \bfv_{\mu}
(\bfw \cdot \del_{\nu} \bfv_{7}) (\bfwt \cdot \del_{\la} \bfv_{N_{+}}) (\bfwt \cdot \del_{\rho} \bfv_{P_{+}}) (\bfwt \cdot \del_{\si} \bfv_{Q_{+}}) \ep^{M_{+} N_{+} P_{+} Q_{+}}. \nonumber \\
\eeqn

Similarly, We compute the contribution to the one-loop effective action by the fundamental fermions,
\beq
\Ga_{\rm 1-loop}^{fund} = - \frac{i}{2} \sum_{f=1}^{n_{f}} \frac{1 + \Ga_{11}}{2}
 \frac{1 + (-)^{|fund|} \ga}{2} \ln \slash{D}_{(f)},
\eeq
where $\slash{D}_{(f)} = \Ga^{\mu} (\del_{\mu} + i v_{\mu})
 + \Ga^{7} i (v_{7} + m_{f}) + \Ga^{M_{+}} i v_{(f)}{}_{M_{+}}$ and $(-)^{|fund|}=-1$.
We take the variation with respect to the gauge field and pick up the relevant terms,
\beqn
\lefteqn{\de \Ga_{\rm 1-loop}^{fund}} \nonumber \\
&=& \frac{1}{16 \pi^{2}} (-)^{|fund|} \sum_{f=1}^{n_{f}} \sum_{\bfw \in \bfW_{fund}} \int d^{5}x sgn( \bfw \cdot \bfv_{7} + m_{f})
\ep^{\mu \nu \la \rho \si} \bfw \cdot \de \bfv_{\mu} (\bfw \cdot \del_{\nu} \bfv_{\la}) (\bfw \cdot \del_{\rho} \bfv_{\si}), \nonumber \\
\eeqn
where $\bfW_{fund} \equiv \{\{ \pm e_{i} \}\}$.

Finally, we obtain
\beqn
\lefteqn{\Ga_{\rm 1-loop}^{fund}} \nonumber \\
&=& \frac{1}{48 \pi^{2}} (-)^{|fund|} \sum_{f=1}^{n_{f}} \sum_{\bfw \in \bfW_{fund}} \int d^{5}x sgn( \bfw \cdot \bfv_{7} + m_{f})
\ep^{\mu \nu \la \rho \si} \bfw \cdot \bfv_{\mu} (\bfw \cdot \del_{\nu} \bfv_{\la}) (\bfw \cdot \del_{\rho} \bfv_{\si}). \nonumber \\
\eeqn

\subsection{Summary of Our Results}
We have exhibited the anomalous interaction on the new phase,
\beqn
\Ga_{\rm 1-loop} &=& \Ga_{\rm 1-loop}^{adj+asym} + \Ga_{\rm 1-loop}^{fund}, \nonumber \\
\Ga_{\rm 1-loop}^{adj+asym} &=& - \frac{i}{2} \Tr \frac{1 + \Ga_{11}}{2} \hat{\rho}_{f \mp} \ln \slash{D}, \nonumber \\
\Ga_{\rm 1-loop}^{fund} &=& - \frac{i}{2} \sum_{f=1}^{n_{f}} \frac{1 + \Ga_{11}}{2} \frac{1 - \ga}{2} \ln \slash{D}_{(f)},
\eeqn
where 
\beqn
\slash{D} &=& \Ga^{\mu} (\del_{\mu} + i v_{\mu}) + \Ga^{7} i v_{7} + \Ga^{M_{+}} i v_{M_{+}}, \nonumber \\
\slash{D}_{(f)} &=& \Ga^{\mu} (\del_{\mu} + i v_{\mu}) + \Ga^{7} i (v_{7} + m_{f}).
\eeqn

We have obtained
\newpage
\beqn
\lefteqn{\Ga_{\rm 1-loop}} \nonumber \\
&=& \frac{1}{48 \pi^{2}} \sum_{\bfw \in \{\{ \pm 2e_{i} \}\} } \int d^{5}x sgn(\bfw \cdot \bfv_{7})
\ep^{\mu \nu \la \rho \si} \bfw \cdot \bfv_{\mu} (\bfw \cdot \del_{\nu} \bfv_{\la}) (\bfw \cdot \del_{\rho} \bfv_{\si}) \nonumber \\
&+& \frac{-1}{48 \pi^{2}} \sum_{f=1}^{n_{f}} \sum_{\bfw \in \bfW_{fund}} \int d^{5}x sgn( \bfw \cdot \bfv_{7} + m_{f})
\ep^{\mu \nu \la \rho \si} \bfw \cdot \bfv_{\mu} (\bfw \cdot \del_{\nu} \bfv_{\la}) (\bfw \cdot \del_{\rho} \bfv_{\si}) \nonumber \\
&+& \frac{-1}{256 \pi^{2}} \sum_{\bfr = adj, asym} \sum_{\bfw \in \bfW_{\bfr}} \int d^{5}x \left[ (\bfw \cdot \bfv_{7})^{2} + \sum_{M_{+}} (\bfwt \cdot \bfv_{M_{+}})^{2} \right]^{- \frac{5}{2}} \bfw \cdot \bfv_{7} \nonumber \\
&& \times \ep^{\mu \nu \la \rho \si} \bfw \cdot \bfv_{\mu}
(\bfwt \cdot \del_{\nu} \bfv_{M_{+}}) (\bfwt \cdot \del_{\la} \bfv_{N_{+}}) (\bfwt \cdot \del_{\rho} \bfv_{P_{+}}) (\bfwt \cdot \del_{\si} \bfv_{Q_{+}}) \ep^{M_{+} N_{+} P_{+} Q_{+}} \nonumber \\
&+& \frac{-1}{64 \pi^{2}} \sum_{\bfr = adj, asym} \sum_{\bfw \in \bfW_{\bfr}} \int d^{5}x \left[ (\bfw \cdot \bfv_{7})^{2} + \sum_{M_{+}} (\bfwt \cdot \bfv_{M_{+}})^{2} \right]^{- \frac{5}{2}} \bfwt \cdot \bfv_{M_{+}} \nonumber \\
&& \times \ep^{\mu \nu \la \rho \si} \bfw \cdot \bfv_{\mu}
(\bfw \cdot \del_{\nu} \bfv_{7}) (\bfwt \cdot \del_{\la} \bfv_{N_{+}}) (\bfwt \cdot \del_{\rho} \bfv_{P_{+}}) (\bfwt \cdot \del_{\si} \bfv_{Q_{+}}) \ep^{M_{+} N_{+} P_{+} Q_{+}}, \nonumber \\
\labeleq{result}
\eeqn
where $\bfW_{fund} \equiv \{\{ \pm e_{i} \}\}$.

The first and the second terms have been computed in \cite{Seiberg, IMS}.
The third and the fourth terms are the anomalous interactions we have found.
These interactions represent a generalized Lorentz force among D4-branes in the multiprobe picture~\cite{Tsey}.


\newpage

\begin{thebibliography}{99}


\bibitem{Seiberg}
N. Seiberg, {\sl Phys. Lett.} {\bf B388} (1996) 753.

\bibitem{IMS}
K. Intriligator, D.R. Morrison and N. Seiberg, {\sl Nucl. Phys.} {\bf B497} (1997) 56.


\bibitem{Tsey}
A. A. Tseytlin and K. Zarembo,
 hep-th/9911246;
 O. Ganor and L. Motl, {\sl JHEP} {\bf 05} (1998) 009,
 hep-th/9803108;
 K. Intriligator, hep-th/0001205.

\bibitem{Th}
C. Boulahouache and G. Thompson, {\sl Int. J. Mod. Phys.} {\bf A13} (1998) 5409,
 hep-th/9801083.

\bibitem{Li}
M. Li, {\sl Nucl. Phys.} {\bf B460} (1996) 351,
 hep-th/9510161;
 M. R. Dougloas, hep-th/9512077;
 M. B. Green, J. A. Harvey and G. Moore, {\sl Class. Quant. Grav.} {\bf 14} (1997) 47,
 hep-th/9605033.

\bibitem{IT}
H. Itoyama and A. Tokura, {\sl Prog. Theor. Phys.} {\bf 99} (1998) 129,
 hep-th/9708123;
 H. Itoyama and A. Tokura, {\sl Phys. Rev.} {\bf D58} (1998) 026002,
 hep-th/9801084.

\bibitem{ITsCMK}
H. Itoyama and A. Tsuchiya, {\sl Prog. Theo. Phys.} {\bf 101} (1999) 1371,
 hep-th/9812177;
 H. Itoyama and T. Matsuo, {\sl Phys. Lett.} {\bf B439} (1998) 46,
 hep-th/9806139;
 B. Chen, H. Itoyama and H. Kihara, {\sl Mod. Phys. Lett.} {\bf A14} (1999) 869,
 hep-th/9810237;
 B. Chen, H. Itoyama and H. Kihara,
 hep-th/9909075.

\bibitem{ITs}
H. Itoyama and A. Tsuchiya, {\sl Prog. Theor. Phys. Suppl.} {\bf 134} (1999) 18,
 hep-th/9904018.


\bibitem{IKKT}
N. Ishibashi, H. Kawai, Y. Kitazawa and A. Tsuchiya, {\sl Nucl. Phys.} {\bf B498} (1997) 467,
 hep-th/9612115.

\end{thebibliography}
\end{document}